# Exploring gender differences on general and specific computer self-efficacy in mobile learning adoption

Yukun Bao[*], Tao Xiong, Zhongyi Hu, Mboni Kibelloh

Center for Modern Information Management,

School of Management, Huazhong University of Science and Technology, Wuhan,

P.R.China, 430074

**Abstract**

Reasons for contradictory findings regarding the gender moderate effect on computer self-efficacy in the adoption of e-learning/mobile learning are limited. Recognizing the multilevel nature of the computer self-efficacy (CSE), this study attempts to explore gender differences in the adoption of mobile learning, by extending the Technology Acceptance Model (TAM) with general and specific CSE. Data collected from 137 university students were tested against the research model using the structural equation modeling approach. The results suggest that there are significant gender differences in perceptions of general CSE, perceived ease of use and behavioral intention to use but no significant differences in specific CSE, perceived usefulness. Additionally, the findings reveal that specific CSE is more salient than general CSE in influencing perceived ease of use while general CSE seems to be the salient factor on perceived usefulness for both female and male combined. Moreover, general CSE was salient to determine the behavioral intention to use indirectly for female despite lower perception of general CSE than male's, and specific CSE exhibited stronger indirect effect on behavioral intention to use than general CSE for female despite similar perception of specific CSE as males'. These findings

---
[*] Corresponding author: Tel: +86-27-87558579; fax: +86-27-87556437.
Email: yukunbao@mail.hust.edu.cn or y.bao@ieee.org

**Gender differences on general and specific computer self-efficacy**

provide important implications for mobile learning adoption and usage.

**Gender differences on general and specific computer self-efficacy**

# 1. Introduction

Gender differences have been examined in various studies regarding the moderate effect in the context of e-learning adoption (e.g. Gefen & Straub, 1997; Ong & Lai, 2006; Barker & Aspray 2006; Kay 2008; Kay 2009; Wang, Wu, & Wang, 2009, Terzis & Economides, 2011). These studies clearly demonstrate that providing more detailed information about gender differences is increasingly important for teachers and learning technology providers. By understanding better gender differences in students' attitudes towards computers, teachers will know how to encourage and improve learning processes for students according to gender. Conceivably, gender differences also play an important role in mobile learning even though it is a relatively new technology. This implies that efforts should be made to fully examine gender differences in mobile learning. However, previous research regarding gender differences in perceptions and acceptance of e-learning learning systems found mixed results. Some studies toward e-learning usage found that males had significantly higher positive perceptions regarding e-learning than females (e.g. Enoch & Soker, 2006; Hoskins & Van Hooff, 2005; Ong & Lai, 2006; Zhou & Xu, 2007). Other studies showed no gender gap regarding perceptions (e.g. Davis & Davis, 2007; Zhang, 2005).

Further studies regarding the gender moderate effect on the explanatory variables that affect technology acceptance also demonstrate contradictory findings. One such variable, examined in this research, is computer self-efficacy (CSE), which defined as an individual judgment of one's capability to use a computer, has played a significant role in an individual's decision to use computers, as well as in the ease with which many of the skills associated with effective computer use are required (Marakas, Yi and Johnson, 1998). Durndell & Thomson (1997) and Ong & Lai, (2006) supported that males' rating of computer self-efficacy is higher than females. However, a few studies could not confirm the gender gap in computer self-efficacy as, for example, a study by Holcomb, King, and Brown (2004) who investigated a sample of students in a distance learning program, and same findings have been found in Imhof, Vollmeyer, & Beierlein (2007). From a logical perspective, the conflicting results suggest the construct of computer self-efficacy under investigation may be unrelated to a specified task and thus be subjected to inappropriate measurements.

In previous studies on information systems (IS) acceptance, CSE has been regarded as a significant predictor of usage intention (Munro et al., 1997; Compeau et al., 1999; Ong & Lai, 2006) as well as has significant effects on other determinants of IS acceptance such as perceived usefulness, perceived ease of use and computer anxiety (Venkatesh, 2000). However, there have been an increasing number of studies which indicated the inconsistence in the relationship between CSE and related variables. For instance, a significant relationship was found between computer self-efficacy and perceived ease of use but none was found between perceived usefulness and information technology (IT) usage (Igbaria and Iivari, 1995). Lopez and Manson (1997) found that CSE had a positive impact on perceived usefulness. Chau (2001) found that computer self-efficacy had a relatively small, but negative, effect on perceived usefulness and no significant effect on perceived ease of use. A study by Ramayah and Aafaqi (2004) on the impact of CSE and cognitive beliefs toward technology adoption in e-library usage among students in a Malaysian university concurred that CSE has a significant direct impact on perceived usefulness and perceived ease of use in e-library usage but no direct impact on usage. Toward the equivocal or contradictory results with regard to the construct of computer self-efficacy, Marakas, Yi and Johnson (1998) suggested that it may be attributable to a general lack off attention to the dynamic, multileveled, and multifaceted nature of it. They proposed that CSE is a multidimensional construct that can be operationalized at both the general computing behavior level and at the specific computer task or application level. General CSE refers to an individual's judgment of efficacy across multiple computer application domains and has been operationalized in many studies in IS acceptance, while application or task specific computer self-efficacy (specific CSE) refers to an individual's perception of efficacy in performing specific computer-related tasks within the domain of general computing (Marakas et al. 1998). Following the seminal work of Marakas, Yi and Johnson, researchers have further distinguished between general CSE and specific CSE (Agarwal et al., 2000; Hsu and Chiu, 2004;Hasan, 2006; Marakas, Johnson and Clay, 2007;Tzeng, 2009). All of the studies argue the need for further research to explore fully the role of general CSE and specific CSE in various areas such as mobile learning.

This paper describes the first study in a program of research aimed at understanding the impact of both general CSE and specific CSE on individual use intention to mobile learning as well as fully exploring and highlighting potential gender differences in perceptions on general CSE and specific CSE, and acceptance in the context of mobile learning. For this purpose, a research model, built upon the Technology Acceptance Model (TAM) by Davis, Bagozzi, and Warshaw (1989), is proposed to include general CSE, specific CSE, perceived usefulness and perceived ease of use. Additionally, the study involves the development of a measure for specific CSE of mobile learning system and a test of its reliability and validity.

The rest of the paper is organized as follows. Section 2 presents our research model and hypotheses, while Section 3 proposes the measurement method and scales. We present the research results in Section 4, followed by discussion in Section 5. Finally, the implications and conclusions of this work are presented in Sections 6.

## 2. Theoretical rational and hypotheses

**Figure 1** shows the research model which was modified from the original TAM developed by Davis (1989). General CSE and specific CSE are added into the research model. Gender is regarded as a moderator variable. The model hypothesizes that gender has a moderating effect on the following relationships: general/specific CSE and perceived ease of use (PEOU); general/specific CSE and perceived usefulness (PU); PEOU and PU; PEOU and behavioral intention to use (BI); and PU and BI.

<center>**<Insert Fig.1 here>**</center>

*2.1 General / specific computer self-efficacy and TAM variables*

General CSE is more a product of a lifetime of related experiences and tends to more closely conform to the definition of computer self-efficacy that is often offered and tested in the IS literature (Carlson and Grabowski 1992, Martocchio 1994). Bandura's (1986) suggested that evaluations of self-efficacy tailored to specific domains provide better predictors of behavior than omnibus assessments. Thus, specific CSE is a state-oriented efficacy that is easier to influence and manipulate (Hsu & Chiu, 2004). Despite the importance of specific CSE to various IS-related outcomes, however, it is only recently that it has begun to attract attention to IS research. For instance, Hsu and Chiu (2004) examined the impact of general Internet CSE and Web-specific CSE on behavioral intentions and actual usage of an Internet-based application. Their results revealed that Web CSE had positive effects on usage intention and actual system usage. In contrast, general Internet CSE demonstrated indirect effects on behavioral intention and usage through its direct effect on attitudes. Additionally, it is worth noting that empirical studies examining the differences between general CSE and specific CSE always conflict in related literature. Johnson and Marakas (2000) found that both levels of CSE (general CSE and specific CSE) have positive effects on computer learning performance. However, Yi & Im (2004) indicated that specific CSE demonstrated a non-significant effect on computer learning performance. Thus, a further study is appealed to explore the relationship among the general CSE, specific CSE, PU and PEOU. Thus, we propose the following hypotheses:

H1a: General CSE will have a positive effect on PU.

H1b: General CSE will have a positive effect on PEOU.

H2a: Specific CSE will have a positive effect on PU.

H2b: Specific CSE will have a positive effect on PEOU.

*2.2 General / specific computer self-efficacy and Gender*

Furthermore, studies have found that males, compared to females, tend to hold higher perceptions of self-efficacy (Busch, 1995; Whitley, 1997, Ong & Lai, 2006). There is evidence to support the view that women show a relatively high tendency toward emotion (Fisk & Stevens, 1993). Venkatesh and Morris (2000) argued that women typically display lower self-efficacy (SE), lower computer aptitude and higher computer anxiety than men. A higher level of anxiety can lead to a lower level of SE and subsequently influence outcome expectations (Compeau & Higgins, 1995). In the context of TAM, Chau (2001) argued that PU reflects a person's beliefs or expectations about outcomes. Hence CSE may be an important factor affecting PU. Therefore, a lower level of CSE is likely to lead to lower levels of PEOU and PU. Based on these arguments, we propose the following hypotheses:

H3a: Male's rating of general CSE is higher than female's.

H3b: Male's rating of specific CSE is higher than female's.

H4a: General CSE will influence PU more strongly for females than males.

H4b: General CSE will influence PEOU more strongly for females than males.

H5a: Specific CSE will influence PU more strongly for females than males.

H5b: Specific CSE will influence PEOU more strongly for females than males.

*2.3 General and specific computer self-efficacy*

As for the relationship between general CSE and specific CSE, Agarwal et al. (2000) pointed out that, although the argument concerning general and specific self-efficacy is theoretically convincing, research on the relationship between the two types of self-efficacy is limited. In the realm of computing, studies have shown that general CSE had a significant effect on specific CSE; e.g. Agarwal et al. (2000) found that general CSE had a significant effect on Windows 95 CSE; however, they also showed that general CSE had a non-significant effect on Lotus 123 efficacy. More recently, Hsu and Chiu (2004) found that general Internet CSE had a significant effect on web-specific CSE. In this study, we propose the hypotheses on the relationship between general CSE and specific CSE and the gender effect as:

H6. General CSE will have a positive effect on specific CSE.

H7. General CSE will influence specific CSE more strongly for females than males.

*2.4 Perceived usefulness*

Perceived usefulness (PU) represents another key determinant of IS acceptance and use. PU refers to the degree to which a person believes that using a system would enhance his or her job performance (Davis, 1989). This implies that mobile learning with a high level of perceived usefulness is one for which a user believes that there is a positive user-performance relationship. A significant body of prior research has shown that perceived usefulness has appositive effect on behavioral intention to use (Venkatesh, 1999, 2000; Venkatesh & Davis, 1996, 2000; Venkatesh & Morris, 2000). According to Venkatesh and Morris (2000), males tend to exhibit stronger and more sensitive attitudes toward task-oriented and instrumental applications of IT than females. Mobile learning is such kind of activity and there is reason to believe that the influence of PU on BI is stronger for males than for females. We hypothesize that:

H8. Male's rating of PU is higher than female's.

H9. PU influences BI more strongly for male than for female.

*2.4 Perceived ease of use*

Davis (1989) referred to perceived ease of use as "the degree to which a person believes that using a particular system would be free of effort. Previous research has shown that men's rating of self-efficacy/computer self-efficacy is higher than women's (Comber et al., 1997; Durndell et al., 2000; Durndell & Hagg, 2002; Whitely, 1997). Venkatesh and Morris (2000) suggested that females have lower SE, lower computer aptitude and higher computer anxiety compared to men and proposed that the influence of perceived ease of use on the intention to use IT is stronger for females than for males. On the other hand, Ong and Lai's (2006) findings suggested that PEOU is more salient in determining PU for females than males. In mobile learning literature, Wang, Wu, and Wang (2009) stated that there are several challenges facing m-learning, such as connectivity, limited processing power and reduced input capabilities. Maniar, Bennett, Hand, and Allan (2008) suggest that many possible technological restrictions impede m-learning adoption, such as small screen size and poor screen resolution. These studies indicate that learners would be more willing to use mobile learning, if they find that the technology can be easily used. Taken together, we, therefore, proposed the following hypotheses:

H10. Male's rating of PEOU is higher than female's.

H11. PEOU influences PU more strongly for female than for male.

H12. PEOU influences BI more strongly for female than for male.

*2.5 Behavioral intention to use*

In this study, intention to use was used as the dependent variable because of its close link to actual behavior (Hu, Clark, & Ma, 2003; Kiraz & Ozdemir, 2006). Intention to use is an attitude, whereas use is a behavior. Substituting the former for the latter may resolve some of the process with causal concerns that Seddon (1997) has raised. In addition, using intention to use as a dependent variable in this study has practical advantages because access to information on the actual use of technology in schools may be too sensitive and thus discourage participation from schools (Teo, 2011).

A significant body of research has indicated that males are more experienced with and have more positive attitudes about computers than do females (e.g., Durndell & Thomson, 1997; Whitely, 1997). It can be said that men are more willing to use computers and mobile devices in the stage of learning. Furthermore, Sáinz & López-Sáez (2010) investigated gender differences in the computer attitudes and the choice of technology-related occupations in Spain and the results revealed that female students displayed more negative computer attitude than male and also made less intense use of technology and computers than their male counterparts. Thus we hypothesize:

H13. Male's rating of BI is higher than female's.

## 3. Methodology

*3.1 Measures*

For the measurement of the latent variables in the model, multiple items were used, based on previously published scales for the various constructs except for specific CSE. For general CSE, we used the ten items from a widely-used and well-validated instrument (Compeau & Higgins, 1995). Items in this instrument asked subjects to rate their ability to perform unspecified computing task using unidentified software. Responses were recorded on a 10-point interval scale starting with 1 (not at all confident) and ending with 10 (totally confident). Specific CSE was measured by six self-developed items, starting with the same stem, "I believe I have the ability to …", followed by the actual tasks within the mobile learning domain. Responses to items on this instrument were recorded on a seven-point Likert-type scale ranging from 1 (strongly disagree) to 7 (strongly agree). As for perceived usefulness, perceived ease of use, and behavioral intention to use, validated items were adapted from prior studies (Davis, 1989; Venkatesh & Davis, 1996; Ong & Lai, 2006). The respondents indicated their agreement or disagreement with the survey instruments using a seven-point Likert-type scale. Similarly, the items were modified to be relevant to the mobile learning usage context. Because the original items were in English, a back-to-back translation procedure was conducted to make the Chinese wording of all the items clear and understandable. A pilot test of 30 subjects was then conducted to further test the instruments. The results of the pilot test were evaluated by using Cronbach's reliability and factor analysis, and all the items showed good convergent and discriminate validities. Appendix presents a list of the items used in this study.

*3.2 Subjects*

The participants for this questionnaire survey consisted of a total of 151 business undergraduate enrolled in an introductory computer science course. Verbal explanation of the purpose and the significance of the study were given. A demonstration of Hub, a mobile learning system, was presented which explained the functions and features of the system. All returned questionnaires were manually checked to ensure there were no missing or ambiguous answers. A total of 143 questionnaires were received, of which 6 were incomplete and were therefore excluded from the final analysis. Among the 137 usable responses, there were 59 females (43%) and 78 males (57%). The average age of the students was 18.7 (Standard deviation (SD) =1.07).

## 4. Data analysis and Results

Following the two-step approach recommend by Anderson and Gerbing (1988), the measurement model was examined and the construct reliability and validity were tested firstly. Then, LISREL 8.72 was used to test the proposed model and the corresponding hypotheses.

### 4.1 Measurement model

The reliability of the five constructs of the proposed research model was examined using Cronbach's alpha, including 0.911 for general CSE, 0.922 for specific CSE, 0.871 for PU, 0.893 for PEOU and 0.861 for BI. All were above 0.8 and exceeded the common threshold value recommended by Nunnally (1978), which indicated high consistency among these constructs.

A confirmatory factor analysis (CFA) was conducted to test the measurement model. Ten common model-fit indices were used to assess the model's overall goodness of fit. As shown in Table 1, all the model-fit indices exceeded their respective common acceptable levels suggested in Gefen, Straub &

Boudreau (2000), which demonstrated a good fit between the model and data.

Reliability and convergent validity of the factors were estimated by composite reliability (CR) and average variance extracted (AVE) (see Table 2). Composite reliability for all the factors in the measure model was above 0.9. The average variance extracted (AVE) for every construct was well above 0.6, indicating good convergent validities (Baggozi & Yi, 1988).

To examine discriminate validity, this study compared the shared variance between constructs with the AVE of each construct. As shown in Table 2, the shared variance between constructs were lower than the AVE of each construct, suggesting good discriminant validity.

**\<Insert Table.1 here\>**

**\<Insert Table.2 here\>**

*4.2 Evaluation of the structural model*

A similar set of model-fit indices was used to examine the structural model (see Table 1). All the ten model-fit measures of the proposed structural model were the same as those of the measure model, indicating a good model fit.

*4.3 Hypothesis testing*

The effects of gender upon general CSE, specific CSE, PU, PEOU, and BI were examined using ANOVAs. Details on mean scores, standard deviation and significant *F* ratios are shown in Table 3. Significant gender differences were founded for general CSE, PEOU, and BI, indicating that male's ratings of general computer self-efficacy, perceived ease of use, and behavioral intention to use mobile learning were higher than female's. Hence H3a, H10, and H13 were supported. However, there was no significant difference in the perception of specific CSE and PU between genders, thus H3b and H8 were not confirmed.

The statistical significance of the relations and direct and indirect effect in the structural model were examined with the data from entire data sample and each of the gender groups. Properties of the causal paths, including standardized path coefficients, the significance of difference, and variance explained for behavioral intention to use in the hypothesized model, are presented in Table 4. As expected, H1a, H1b, H2a, H2b and H6 were supported. Compared to male, female exhibited a greater salient effect on general CSE and specific CSE in determining PU (H4a and H5a were supported), and PEOU (H4b and H5b were supported) in addition to placing a greater emphasis on PEOU in determining PU (H11 was supported). However, the hypothesized stronger effects of general CSE on specific CSE and for female did not confirm (H7 was not supported). Furthermore, male weighted PU more strongly in determining BI than female did (H9 was supported). Contrary to H12, the direct effect of PEOU on BI between female and male was not significant. Table 5 summarizes the testing results of hypotheses.

The direct, indirect and total effect of general CSE, specific CSE, perceived usefulness and perceived ease of use on behavioral intention to use are summarized in Table 6. It should be noted that specific CSE has a stronger indirect effect on behavioral intention to use than general CSE for female.

**<Insert Table.3 here>**

**<Insert Table.4 here>**

**<Insert Table.5 here>**

**<Insert Table.6 here>**

## 5. Discussions

The aim of this study is to explore the gender difference in mobile learning adoption. Different form previous research, we divided computer self-efficacy, widely used as one important determinant in IS acceptance research, into both general CSE and specific CSE, and extended them into TAM model. Thus, we can fully explore and highlight the gender difference in perceptions of general CSE, specific CSE and their relationships and direct or indirect effects with perceived usefulness, perceived ease of use and individual use intention to mobile learning.

Regarding the differences in perceptions of general CSE and specific CSE, male's ratings of general CSE is higher than female's as was hypothesized, consistent with Ong and Lai's work (2006). Note that general CSE was used in their study, although it was not indicated. However, the current findings did not confirm male's higher ratings for specific CSE, demonstrating female may be more positive when providing explicit tasks or instructions related to study. This means that female may perceive their perceptions of general CSE and specific CSE differently and perhaps general CSE and specific CSE are different constructs in nature as indicated by Marakas, Yi and Johnson (1998). Another possible reason may be from gender stereotypes. Consistently, girls and boys generally believed males to be better at computing than females (Fisk & Stevens, 1993; Durndell, Glissov & Siann, 1995; Clegg, Mayfield, & Trayhurn, 1999; Arbaugh, 2000; Schott & Selwyn, 2000; Schumacher & Morahan-Martin, 2001). Thus, female may exhibit lower general CSE but be more positive when providing details on computing tasks. To this regard, for sure, well-founded conclusion requires additional research.

Furthermore, we also investigate the relationships and direct or indirect effects between general CSE and specific CSE as well as with perceived usefulness, perceived ease of use and individual use intention in the context of mobile learning, demonstrating how and to which extent these two genders differ from each other. The results support some of our hypothesis. Tables 4 and Table 6 show the statistical significant relationships and direct or indirect effects for both genders. Consistent with the study examining the relationship between general computer self-efficacy and task-specific computer self-efficacy (Marakas, Yi and Johnson, 1998), the present study confirmed that general CSE contributed to the shape of an individual's specific CSE. Interestingly, there is no gender difference in the effect of general CSE on specific CSE. This means that in both male and females; mind and memories are not stored in isolation, but in networks in which each item is linked to many others through connections referred to as associations (Collins & Loftus, 1975). For example, the association formed by ''carrying out the task' and ''using a specific application'' which occur at the same time, further suggesting that a system is essentially related to the task it helps to accomplish.

As we expected, both general CSE and specific CSE appeared to be a significant determinant of perceived usefulness and perceived ease of use for both female and male, indicating users who have higher general CSE or specific CSE are likely to have more positive usefulness and ease of use beliefs. It should be noted that the total effect of specific CSE on perceived ease of use is greater than general CSE both for female and male, while the total effects of general CSE on perceived usefulness are greater than specific CSE for both female and male. This finding suggests both female and male tend to increase their ease of use beliefs when providing clear understanding on what they should deal with and with what kind of skill needed.

Though female's rating of general CSE being lower than male's, their perception of general computer self-efficacy was still a more salient determinant affecting behavioral intention to use in a indirect way through perceived usefulness, perceived ease of use and specific CSE. In addition, specific

CSE had a stronger indirect effect on behavioral intention to use than general CSE, in particular for female. This indicated female's perception of specific CSE was a significantly better predictor for behavioral intention to use mobile learning than general CSE. This lends empirical confirmation to Bandura's (1997) belief that self-efficacy should be measured in a domain by using actual tasks to provide explanatory and predictive capability and also provides evidence on the claims that the more the measure of self-efficacy moves from specific tasks to general measure, the greater the deterioration in the relationship between self-efficacy and its consequences such as usage (Gist, 1987; Marakas, Yi and Johnson, 1998). Because an individual's intention to new technology relies on that person's ability in accomplishing tasks, specific CSE, as a task-oriented instrument, shows considerable promise in mobile learning adoption.

Although this study focused on the gender differences in perceptions of general and specific CSE and their relationships with relative variables, the empirical results also revealed some findings regarding the gender differences in perceptions of perceived usefulness, perceived ease of use and behavioral intention to use as well as relationships among them. A significant gender variation was found in perceptions of perceived ease of use and behavioral intention to use but not for perceived usefulness. This means that both genders perceive mobile learning helpful for their studies regardless the differences in general CSE, perceived ease of use and behavior intention to use. As for the relative influences of perceived usefulness and perceived ease of use on behavior intention to use, male's perception of usefulness was the more significantly direct and more salient than female's in determining behavioral intention to use mobile learning while female weighted more on perceived ease of use for their decisions on taking mobile learning, in line with the results in Ong and Lai (2006).

## 6. Conclusions

The main contributions of this study are three fold. First, this study provides new insight into the nature of computer self-efficacy at general and specific level, and empirical evidence on gender differences in perceptions of both of them in the mobile learning adoption, though the reasons remain open. Second, this research reveals that general CSE can positively influence specific CSE for both female and male, and specific CSE is more salient than general CSE in influencing perceived ease of use while general CSE seems to be the salient factor on perceived usefulness for both female and male. As such, it is of significance and contributable to reveal how the different levels of computer self-efficacy affect the relative determinants in TAM within the context of mobile learning. Third, one rather noticeable result of this study was the differences in how general CSE and specific CSE influenced behavioral intention to use. We found general CSE was salient to determine the behavioral intention to use indirectly for female despite lower perception of general CSE, and specific CSE exhibited stronger indirect effect on behavioral intention to use than general CSE for female despite similar perception of specific CSE as male's. This provides valuable implications for teaching and training practice as to be discussed later.

As with all empirical research, this study has certain limitations. First, investigating the gender differences in perceptions of computer self-efficacy at both general and specific levels in mobile learning is a relatively new topic for IS/IT researchers. The findings and their implications presented here were obtained from a single study within a university. As such, generalization to a general population with difference environment settings, such as in working setting, must be approached with cautions. Second, gender differences regarding mobile learning adoption is quite a complex topic, thus other important variables should be added by the future studies. Third, sample characteristics are an issue here even within academic learning context. The sample was of full-time young undergraduate students. This will limit the extent of the generalization of findings to other students groups, such as part time or mature students. In addition, the sample size has room for improvement.

In our findings, there are several theoretical and practical implications for mobile learning. The theoretical implication is that the importance of a continuing and rigorous investigation into the CSE at both the general and specific levels is of significant value. We believe that such distinguishing manipulation on CSE deserve potentially consequential worth. The empirical evidences attained in this present study offer as yet another step toward understanding the nature of this widely used construct in IS/IT literature. The practical implications come from what have been indicated in female's perceptions of general and specific CSE and their relationships with behavioral intention to use as mentioned above. The training programs should be well designed to improve female's perceptions of computer self-efficacy, in particular the specific CSE. More attention should be focused on task-specific skills when designing the modules for training program or related courses.

## Conclusions

The authors would like to thank the anonymous reviewers for their valuable suggestions and constructive comments. This work was supported by the Fundamental Research Funds for the Central Universities (2012QN208-HUST) and a grant from the Modern Information Management Research Center at Huazhong University of Science and Technology.

**Gender differences on general and specific computer self-efficacy**

**Appendix.**

Questionnaire items

General computer self-efficacy

I could complete the job using the software package…

GCSE1 …if there was no one around to tell me what to do as I go.

GCSE2 …if I had never used a package like it before.

GCSE3 …if I had only the software manuals for reference.

GCSE4 …if I had seen someone else using it before trying it myself.

GCSE5 …if I could call someone for help if I got stuck.

GCSE6 …if someone else had helped me get started.

GCSE7 …if I had a lot of time to complete the job for which the software was provided.

GCSE8 …if I had just the built-in help facility for assistance.

GCSE9 …if someone showed me how to do it first.

GCSE10 …if I had used similar packages before this one to do the same job.

Specific computer self-efficacy (self-developed for this study)
I believe I have the ability to

SCSE1 …configure the hardware to access the mobile learning system through personal mobile devices such as laptop or other PDAs.

SCSE2…install and configure the software to access the mobile learning system through personal mobile devices such as laptop or other PDAs.

SCSE3 …browse and retrieve the learning materials.

SCSE4 …download the learning materials needed.

SCSE5 …upload any files or pictures required.

SCSE6 …posit questions for help from instructors or classmates.

SCSE7 …use the online chatting room.

Perceived usefulness (PU)

PU1 Using the mobile learning system improves my study performance.

PU2 Using the mobile learning system enhances my effectiveness in my study.

PU3 Using the mobile learning system in my study improves my productivity.

PU4 I find the mobile learning system to be useful in my study.

Perceived ease of use (PEOU)

PEOU1 My intention with the mobile learning system is clear and understandable.

PEOU2 Interacting with the mobile learning system does not require a lot of my mental effort.

PEOU3 I find the mobile learning system to be easy to use.

PEOU4 I found it easy to get the mobile learning system to do what I want it to do.

Behavioral intention to use (BI)

BI1 Assuming that I had access to the mobile learning system, I intend to use it.

BI2 Given that I had access to the mobile learning system, I predict that I would use it.

Table 1

Fit indices for measurement and structural models

| Goodness-of-fit measures | Recommended value | Measurement model | Structural model |
|---|---|---|---|
| $\chi^2/df$ | $\leq 3$ | 1.974 | 1.974 |
| GFI | $\geq 0.8$ | 0.88 | 0.88 |
| RMSR | $\leq 0.05$ | 0.036 | 0.036 |
| SRMR | $\leq 0.08$ | 0.052 | 0.052 |
| RMSEA | $\leq 0.08$ | 0.064 | 0.064 |
| NFI | $\geq 0.9$ | 0.97 | 0.97 |
| CFI | $\geq 0.9$ | 0.98 | 0.98 |
| AGFI | $\geq 0.8$ | 0.83 | 0.83 |
| PNFI | $\geq 0.5$ | 0.87 | 0.87 |
| PGFI | $\geq 0.5$ | 0.75 | 0.75 |

Table 2

Reliability, average variance extracted and discriminant validity

| Factor | CR[a] | 1 | 2 | 3 | 4 | 5 |
|---|---|---|---|---|---|---|
| 1.General CSE | 0.922 | **0.827** | | | | |
| 2.Specific CSE | 0.931 | 0.721 | **0.837** | | | |
| 3.Peceived usefulness | 0.942 | 0.733 | 0.688 | **0.854** | | |
| 4.Perceived ease of use | 0.912 | 0.687 | 0.753 | 0.756 | **0.817** | |
| 5.Behavioural intention to use | 0.925 | 0.702 | 0.726 | 0.731 | 0.741 | **0.832** |

Notes: Diagonal elements are the average variance extracted. Off-diagonal elements are the shared variance.

[a]CR, composite reliability.

Table 3

Descriptive statistics and AVOVAs testing results

| | Female (n=59) | | Male (n=78) | | Significance of difference between female and male *F* ratios |
|---|---|---|---|---|---|
| | Mean | SD | Mean | SD | |
| General CSE | 4.97 | 1.14 | 5.66 | 0.96 | 14.23[***] |
| Specific CSE | 5.21 | 1.21 | 5.33 | 1.26 | **ns** |
| PU | 5.16 | 1.17 | 5.28 | 1.02 | **ns** |
| PEOU | 4.67 | 1.06 | 5.21 | 0.94. | 14.21[***] |
| BI | 5.03 | 1.29 | 5.68 | 0.86 | 5.46[*] |

**ns**, not significant;

[*] $P<0.05$;

[**], $P<0.01$;

[***], $P<0.001$.

Table 4

Gender differences in relationships of general/specific CSE-PU, general/specific CSE-PEOU, PU-BI, PEOU-PU and PEOU-BI

| | Entire sample | | Female (n=59) | | Male (n=78) | | difference between female and male |
|---|---|---|---|---|---|---|---|
| | $R^2$ | $\beta$ | $R_f^2$ | $\beta_f$ | $R_m^2$ | $\beta_f$ | |
| BI | 0.44 | | 0.42 | | 0.45 | | *** |
| General CSE-PU | | 0.32*** | | 0.38*** | | 0.29** | *** |
| General CSE-PEOU | | 0.30*** | | 0.35*** | | 0.22** | *** |
| General CSE-specific CSE | | 0.28** | | 0.31*** | | 0.27** | **ns** |
| Specific CSE-PU | | 0.43*** | | 0.51*** | | 0.24** | *** |
| Specific CSE-PEOU | | 0.43*** | | 0.55*** | | 0.34*** | *** |
| PU-BI | | 0.41*** | | 0.29 **ns** | | 0.43*** | *** |
| PEOU-PU | | 0.34*** | | 0.38*** | | 0.25** | *** |
| PEOU-BI | | 0.39*** | | 0.42*** | | 0.38*** | **ns** |

**ns**, not significant;

* P<0.05;

** , P<0.01;

*** , P<0.001.

Table 5

Summary of testing results

|  |  | Hypothesis | Result |
|---|---|---|---|
| Perception |  |  |  |
|  | General CSE | Male>Female | Supported |
|  | Specific CSE | Male>Female | Not supported |
|  | PU | Male>Female | Not supported |
|  | PEOU | Male>Female | Supported |
|  | BI | Male>Female | Supported |
| Relationship |  |  |  |
|  | General CSE-PU | Female>Male | Supported |
|  | General CSE-PEOU | Female>Male | Supported |
|  | General CSE-specific CSE | Female>Male | Not supported |
|  | Specific CSE-PU | Female>Male | Supported |
|  | Specific CSE-PEOU | Female>Male | Supported |
|  | PU-BI | Male>Female | Supported |
|  | PEOU-PU | Female>Male | Supported |
|  | PEOU-BI | Female>Male | Not supported |

Table 6

Gender differences between the direct and indirect effects of general/specific CSE, PU, PEOU and BI

| DV | IV | Direct effect | | | Indirect effect | | | Total effect | | |
|---|---|---|---|---|---|---|---|---|---|---|
| | | ES | F | M | ES | F | M | ES | F | M |
| BI | PU | 0.41 | 0.29 | 0.43 | 0.00 | 0.00 | 0.00 | 0.41*** | 0.29**ns** | 0.43*** |
| | PEOU | 0.39 | 0.42 | 0.38 | 0.14 | 0.11 | 0.11 | 0.53*** | 0.53*** | 0.49*** |
| | General CSE | 0.00 | 0.00 | 0.00 | 0.39 | 0.41 | 0.29 | 0.39*** | 0.41*** | 0.29** |
| | Specific CSE | 0.00 | 0.00 | 0.00 | 0.40 | 0.44 | 0.27 | 0.40*** | 0.44*** | 0.27** |
| PU | PEOU | 0.34 | 0.38 | 0.25 | 0.00 | 0.00 | 0.00 | 0.34** | 0.38*** | 0.25** |
| | General CSE | 0.32 | 0.38 | 0.29 | 0.26 | 0.36 | 0.14 | 0.58*** | 0.74*** | 0.43*** |
| | Specific CSE | 0.43 | 0.51 | 0.24 | 0.15 | 0.21 | 0.09 | 0.57*** | 0.72*** | 0.33** |
| PEOU | General CSE | 0.3 | 0.35 | 0.22 | 0.12 | 0.17 | 0.09 | 0.42*** | 0.52*** | 0.31** |
| | Specific CSE | 0.43 | 0.55 | 0.34 | 0.00 | 0.00 | 0.00 | 0.43*** | 0.55*** | 0.34** |
| Specific CSE | General CSE | 0.28 | 0.31 | 0.27 | 0.00 | 0.00 | 0.00 | 0.28** | 0.31** | 0.27** |

**ns**, not significant;

* $P<0.05$;

**, $P<0.01$;

***, $P<0.001$.

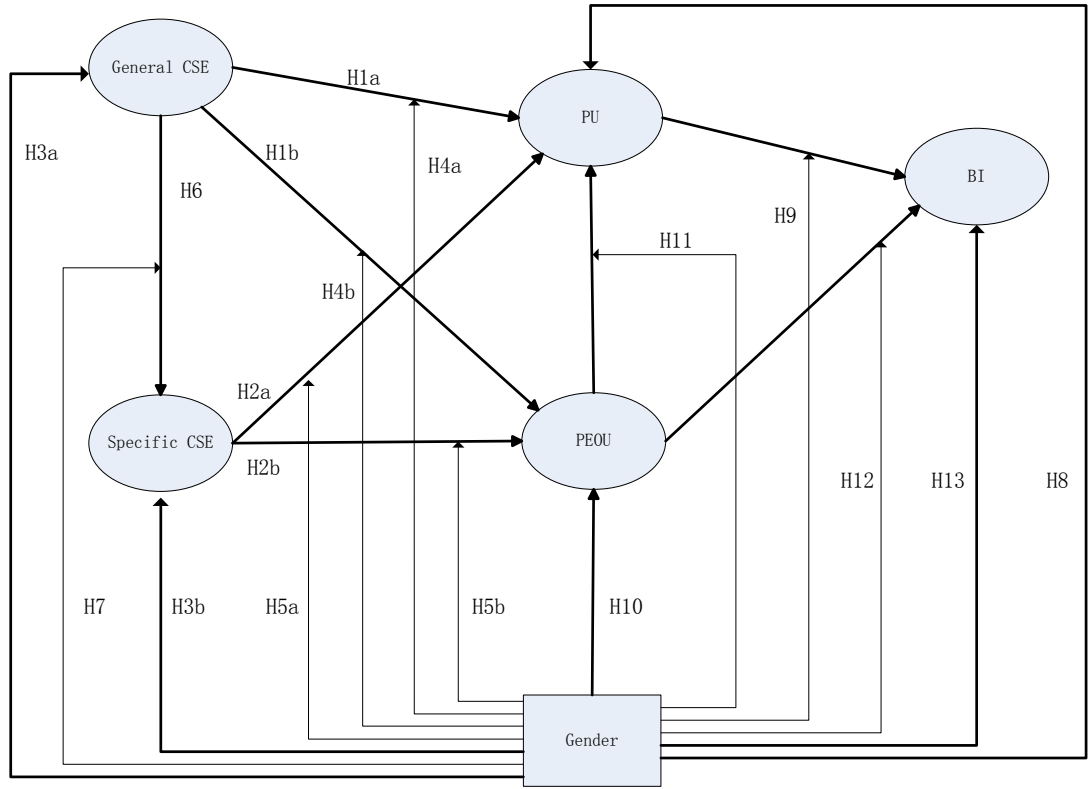

Figure 1 Research model